\documentclass[conference]{IEEEtran}
\IEEEoverridecommandlockouts
\usepackage{amsmath,amssymb,amsfonts}
\usepackage{algorithmic}
\usepackage{graphicx}
\usepackage[dvipsnames]{xcolor}
\usepackage{textcomp}
\def\BibTeX{{\rm B\kern-.05em{\sc i\kern-.025em b}\kern-.08em T\kern-.1667em\lower.7ex\hbox{E}\kern-.125emX}}
    
\usepackage{acronym}
\usepackage{siunitx}
\usepackage{cuted}
\usepackage{epsfig,pstricks}
\usepackage{todonotes}
\usepackage[most]{tcolorbox}
\usepackage[utf8]{inputenc}
\usepackage[T1]{fontenc}
\makeatletter
\let\MYcaption\@makecaption
\makeatother
\usepackage{caption}
\usepackage[font=footnotesize]{subcaption}
\makeatletter
\let\@makecaption\MYcaption
\makeatother

\usepackage{tikz}
\usepackage{pgfplots}
\usepackage{pgfmath}

\definecolor{safeblue}{HTML}{0072B2}
\definecolor{safeorange}{HTML}{D55E00}
\definecolor{safebeige}{HTML}{E69F00}
\definecolor{safecyan}{HTML}{56B4E9}
\definecolor{safegreen}{HTML}{009E73}
\definecolor{safeyellow}{HTML}{F0E442} 
\definecolor{safeplum}{HTML}{CC79A7}
\definecolor{safepurple}{HTML}{332288}

\definecolor{mygray}{HTML}{687681}

\pgfplotsset{%
compat=1.17,
width=\linewidth,
yminorticks=true,
separate axis lines,
axis background/.style={fill=white},
xmajorgrids,
ymajorgrids,
grid style={loosely dotted, black},
tick style={color=black},
label style={font=\footnotesize\rmfamily},
tick label style={font=\small},
legend style={font=\scriptsize\sffamily, cells={anchor=west}, legend plot pos=left, draw=mygray},
legend pos=south east}

\def\clap#1{\hbox to 0pt{\hss#1\hss}}

\usepackage{microtype}

\acrodef{CIR}[CIR]{channel input response}
\acrodef{UL}[UL]{uplink}
\acrodef{DL}[DL]{downlink}
\acrodef{DL-PRS}[DL-PRS]{downlink positioning reference signals}
\acrodef{SRS}[SRS]{sounding reference signals}
\acrodef{TDoA}[TDoA]{time difference of arrival}
\acrodef{ToA}[ToA]{time of arrival}
\acrodef{Multi-RTT}[Multi-RTT]{multi-round trip time}
\acrodef{TDoA}[TDoA]{time difference of arrival}
\acrodef{DL-AOD}[DL-AOD]{downlink angle of departure}
\acrodef{UL-AOA}[UL-AOA]{Uplink Angle of Arrival}
\acrodef{LOS}[LOS]{line of sight}
\acrodef{NLOS}[NLOS]{non-line of sight}
\acrodef{OLOS}[OLOS]{obstructed line of sight}
\acrodef{LPHAP}[LPHAP]{low power high accuracy positioning}
\acrodef{RedCAP}[RedCAP]{reduced capability}
\acrodef{UE}[UE]{user equippment}
\acrodef{TRP}[TRP]{transmission and reception point}
\acrodefplural{TRP}[TRPs]{transmission and reception points}
\acrodef{InF}[InF]{indoor factory scenario}
\acrodef{LCM}[LCM]{life cycle management}
\acrodef{CIR}[CIR]{channel input response}
\acrodef{PDP}[PDP]{power delay profile}
\acrodef{DP}[DP]{delay profile}
\acrodef{RSRP}[RSRP]{reference signal received power}
\acrodef{RSRPP}[RSRPP]{reference signal received path power}
\acrodef{RSTD}[RSTD]{reference signal timing difference}
\acrodef{LMF}[LMF]{location management function}
\acrodef{KPI}[KPI]{key performance indicator}
\acrodefplural{KPI}[KPIs]{key performance indicator}
\acrodef{PRU}[PRU]{positioning reference unit}
\acrodef{gNB}[gNB]{next-generation NodeB}
\acrodef{SNR}[SNR]{signal to noise ratio}
\acrodef{SINR}[SINR]{signal to interference noise ratio}
\acrodef{NR}[NR]{new radio}
\acrodef{RAT}[RAT]{radio access technology}
\acrodef{CDF}[CDF]{cumulative distribution function}
\acrodef{FLOPS}[FLOPS]{floating-point operations per second}
\acrodef{AI}[AI]{artificial intelligence}
\acrodef{ML}[ML]{machine learning}
\acrodef{ARP}[ARP]{antenna radiation point}
\acrodef{RAT}[RAT]{radio access technology}
\acrodef{ATOA}[ATOA]{Additional Time of Arrival}
\acrodef{ACS}[ACS]{Assocuation and Calibration Spot}
\acrodef{IPD}[IPD]{Inter-Point Distance}
\acrodef{IMU}[IMU]{inertial measurement unit}
\usepackage[normalem]{ulem}

\usepackage[tikz]{bclogo}
\usepackage[framemethod=tikz]{mdframed}
\usepackage{lipsum}

\usepackage{awesomebox}

\definecolor{bgblue}{RGB}{245,243,253}
\definecolor{ttblue}{RGB}{91,194,224}

\mdfdefinestyle{mystyle}{%
  rightline=true,
  innerleftmargin=10,
  innerrightmargin=10,
  outerlinewidth=3pt,
  topline=false,
  rightline=true,
  bottomline=false,
  skipabove=\topsep,
  skipbelow=\topsep
}

\newtcolorbox{myboxi}[1][]{
  breakable,
  title=#1,
  colback=white,
  colbacktitle=white,
  coltitle=black,
  fonttitle=\bfseries,
  bottomrule=0pt,
  toprule=0pt,
  leftrule=3pt,
  rightrule=3pt,
  titlerule=0pt,
  arc=0pt,
  outer arc=0pt,
  colframe=black,
}

\newtcolorbox{myboxii}[1][]{
  breakable,
  freelance,
  title=#1,
  colback=white,
  colbacktitle=white,
  coltitle=black,
  fonttitle=\bfseries,
  bottomrule=0pt,
  boxrule=0pt,
  colframe=white,
  overlay unbroken and first={
  \draw[red!75!black,line width=3pt]
    ([xshift=5pt]frame.north west) -- 
    (frame.north west) -- 
    (frame.south west);
  \draw[red!75!black,line width=3pt]
    ([xshift=-5pt]frame.north east) -- 
    (frame.north east) -- 
    (frame.south east);
  },
  overlay unbroken app={
  \draw[red!75!black,line width=3pt,line cap=rect]
    (frame.south west) -- 
    ([xshift=5pt]frame.south west);
  \draw[red!75!black,line width=3pt,line cap=rect]
    (frame.south east) -- 
    ([xshift=-5pt]frame.south east);
  },
  overlay middle and last={
  \draw[red!75!black,line width=3pt]
    (frame.north west) -- 
    (frame.south west);
  \draw[red!75!black,line width=3pt]
    (frame.north east) -- 
    (frame.south east);
  },
  overlay last app={
  \draw[red!75!black,line width=3pt,line cap=rect]
    (frame.south west) --
    ([xshift=5pt]frame.south west);
  \draw[red!75!black,line width=3pt,line cap=rect]
    (frame.south east) --
    ([xshift=-5pt]frame.south east);
  },
}

\begin{document}

\title{5G Positioning Advancements with AI/ML}

\author{
    Mohammad Alawieh,     Georgios Kontes\\
	\textit{Fraunhofer Institute for Integrated Circuits IIS, Nuremberg, Germany}\\
  \texttt{\footnotesize\{mohammad.alawieh, georgios.kontes\}@iis.fraunhofer.de}
}


\maketitle
\begin{abstract}
This paper provides a comprehensive review of AI/ML-based direct positioning within 5G systems, focusing on its potential in challenging scenarios and conditions where conventional methods often fall short. Building upon the insights from the technical report TR38.843, we examine the Life Cycle Management (LCM) with a focus on to the aspects associated direct positioning process. We highlight significant simulation results and key observations from the report on the direct positioning under the various challenging conditions. Additionally, we discuss selected solutions that address measurement reporting, data collection, and model management, emphasizing their importance for advancing direct positioning.
\end{abstract}

\begin{IEEEkeywords}
5G, direct positioning, ML, AI.
\end{IEEEkeywords}

\section{Introduction}\label{sec:introduction}

The 3rd Generation Partnership Project (3GPP) has played a pivotal role in advancing 5G positioning technologies through its previous releases. 3GPP Release 16 \cite{tr38855} laid the foundational groundwork by defining both~\ac{UL} and~\ac{DL} procedures and incorporating signal designs to support the identified positioning use cases. Building on this foundation, Release 17 \cite{tr38857} brought significant enhancements, introducing signalling to counteract timing errors and enhance directional estimation methods. Additionally, this release enabled positioning when a device is in an inactive state and reduced latency positioning. This release further introduced techniques to improve positioning accuracy in~\ac{NLOS} scenarios.  Release 18 introduced sidelink positioning, carrier phase and carrier aggregation to further enhance the positioning accuracy. These advancements have paved the way for numerous applications and use cases, but they also highlight the inherent challenges, particularly in in~\ac{NLOS} and complex in~\ac{LOS} environments where traditional methods might not always yield the desired accuracy.

In response to the evolving requirements and complexities in 5G positioning technologies and the persistent challenges in NLOS and complex LOS scenarios, Release 18 initiated an AI/ML study item to explore the potential of AI and machine learning in the context of positioning. This study,  as provided in technical report TR38.843  \cite{tr38843} , identified two primary modalities through which AI/ML could be beneficial:

\begin{itemize}
    \item Direct AI/ML Positioning: This is where the AI/ML model directly outputs the location of the~\ac{UE}. A typical example is the use of fingerprinting, which employs channel observations as the AI/ML model's input to ascertain the~\ac{UE}'s location. 
    
    \textit{It's worth to clarify a prevalent misunderstanding: while fingerprinting falls under the umbrella of direct positioning, they aren't synonymous. Fingerprinting is  one of several techniques within the broader definition of direct positioning.}
    
    \item AI/ML Assisted Positioning: Rather than directly determining the~\ac{UE}'s location, here the AI/ML model aids conventional approaches by outputting  measurements or by refining existing measurements. For instance, the model identifies y~\ac{LOS} versus~\ac{NLOS} conditions, or provides insights into the timing or angle of measurements and even the likelihood of such measurements.  In some of these challenging environments, AI/ML  complements  traditional methods by providing intermediate AI/ML measurements to assist the classical positioning techniques.
\end{itemize}
\begin{figure}[!htp]
    \centering
    \includegraphics[width=1\columnwidth] {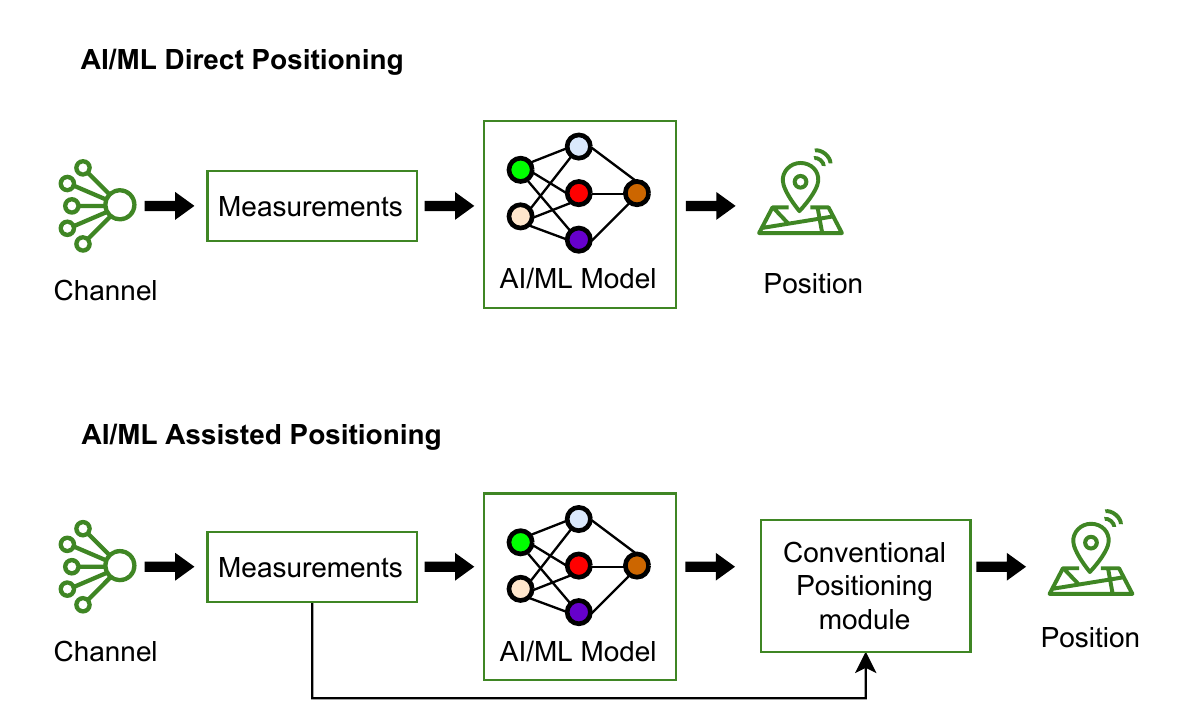}
    \caption{Example realizations on AI/ML direct and assisted positioning.  }
    \label{fig:directassisted}
\end{figure}



Direct positioning emerges as a crucial solution in this context. Serving either as a complementary tool or even an alternative to the classical methods of Release 16 by utilizing  the capabilities of AI/ML in~\ac{NLOS} conditions. In scenarios where classical approaches down-perform with uncertainties, direct positioning stands out by potentially bridging the performance gap.

\renewcommand\bcStyleTitre[1]{\large\textcolor{OliveGreen}{#1}}
\begin{bclogo}[
  couleur=White, 
  arrondi=0,
  logo=\bclampe,
  barre=none,
  noborder=true]{\itshape \textbf{Direct Positioning, likely for future 5G releases, promises effective performance in challenging environments without extensive network adjustments.}}
\end{bclogo}

In the upcoming sections, we'll take a closer look at 5G positioning and focus on direct positioning with AI/ML. Section \ref{sec:challenges} examines specific scenarios where direct positioning is vital, along with the challenges and requirements it entails. Section \ref{sec:3gpp_sota} discusses Life Cycle Management (LCM) to support these requirements. Section \ref{sec:eval} presents findings from the TR38.843 report, assessing the accuracy of direct positioning under different conditions. Lastly, in the last section, we discuss solutions for challenges in measurement reporting, data collection, and model management. These sections collectively provide a comprehensive overview of AI/ML-based direct positioning.

\section{ Direct Positioning Requirements and Anticipated Challenges }\label{sec:challenges}

According to TS 22.261\cite{ts22261}, 5G positioning services have been categorized into various service levels, each characterized by different metrics such as accuracy, positioning service availability, latency, and energy consumption. Providing a simplified overview, the identified absolute horizontal accuracy ranges between 30 cm and 3 m. Furthermore, this accuracy should be available at least 95\% of the time for both indoor and outdoor scenarios.

\begin{figure}[!htbp]
    \centering
    \includegraphics[width=0.45\textwidth] {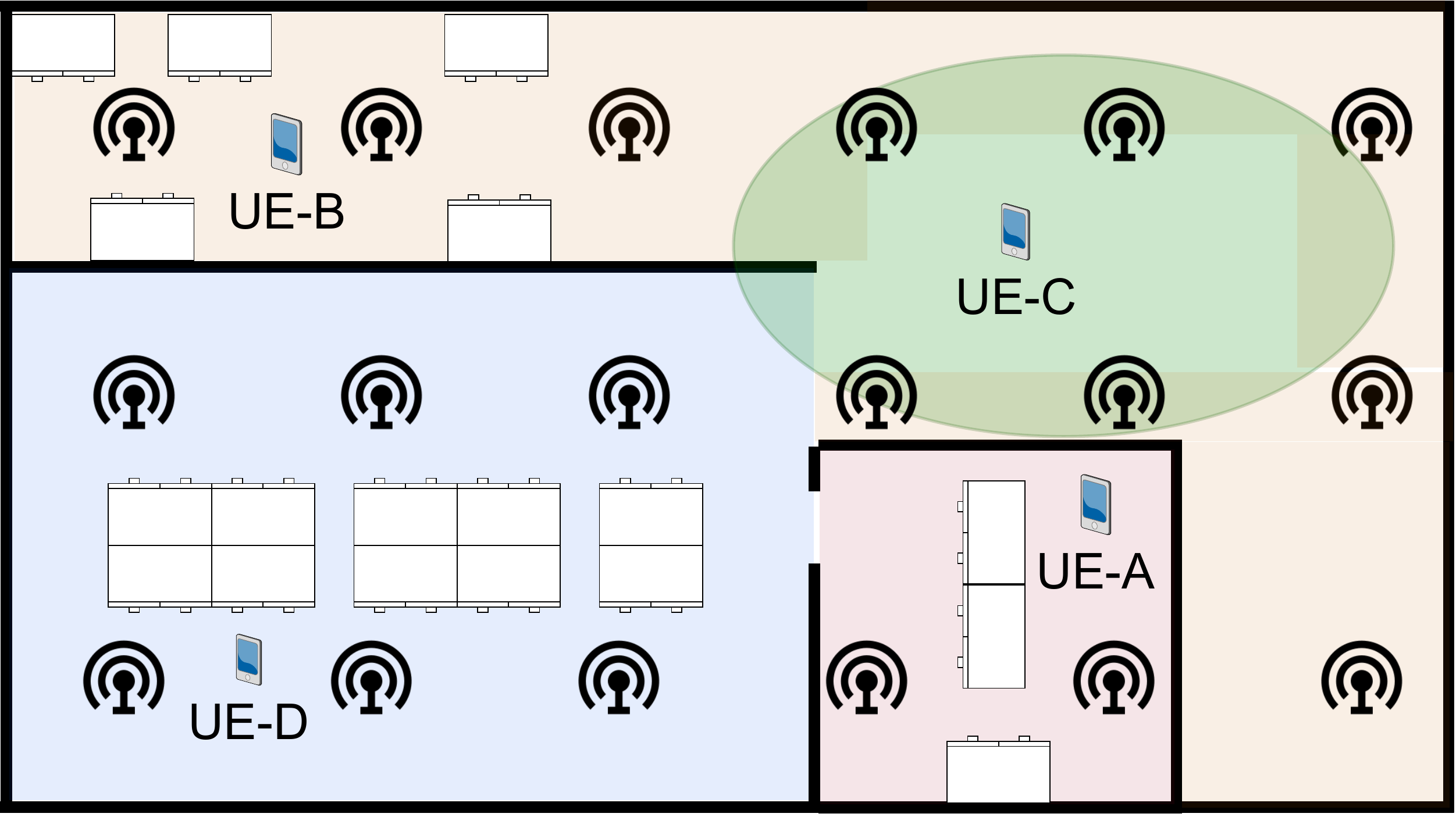}
    \caption{Illustrative representation of a hybrid indoor factory environment, emphasizing different reception conditions across different zones (UE-A to UE-D). }
    \label{fig:InFexample}
\end{figure}

In the positioning evaluations of Release 17, the stringent accuracy requirements are met, but only under specific conditions and configurations. In Fig.~\ref{fig:InFexample}, we illustrate an example of an~\ac{InF}. Echoing the positioning evaluations presented in \cite{tr38843}, this scenario features 18~\acp{TRP}. Specific obstructions, clutter, and diverse reception conditions segment the environment into areas associated with UE-A, UE-B, UE-C, and UE-D. Within UE-A's vicinity, the environment is predominantly characterized by ~\ac{NLOS} conditions, posing significant challenges for traditional positioning methods and thus underscoring the potential of direct positioning approach. In contrast, UE-C is enveloped in an ~\ac{LOS} setting, suggesting that conventional, non-ML positioning methods would provide an optimal performance. UE-B is situated in a scenario exhibiting challenging LOS conditions, caused by near reflections and unfavourable geometry. In such complex environments, as experienced by UE-B, AI/ML-assisted positioning could potentially bolster the accuracy and reliability of established techniques. Lastly, UE-D's location is dominated by heavy~\ac{OLOS} due to dense clutter; evaluations in \cite{ECpaper} suggest that conventional positioning approaches falter in such conditions, hinting at the relevance of direct positioning for these scenarios.

Direct positioning within 5G systems promises enhanced location accuracy for scenarios such as for UE-A and UE-D. However, its effective implementation and integration in the 5G AI/ML ~\ac{LCM} faces a multitude of challenges related to both the complexity of the 5G system and the operation requirements and constraints of the direct positioning.

\begin{itemize}
  \item \textbf{Functionality Across Varied Deployment Scenarios}: 
   Traditional 5G positioning techniques often rely on specific conditions: synchronization, a high number of \acp{TRP}, calibration of antennas and delays, and precise determination of the \ac{ARP}. However, these conditions might not always be present or feasible in every scenario.
 Consequently,  for direct positioning to be effectively integrated into 5G systems, it must be adaptable to a range of deployment considerations.  Hence, direct positioning needs to provide operational flexibility which can adjust to different conditions, although its performance might vary based on the presence and accuracy of these considerations.
  
    \item \textbf{Vendor Compatibility}: The 5G standard does not incorporate the variety of vendor solutions, which may have different realizations. For instance, considering model inference at the UE, performance variations among different device types can be expected due to different constraints in terms of complexity.  Conversely,  in the context of network-sided models, the exact method used to extract measurements from received channels at either the \ac{TRP} or \ac{UE} might not be transparent to the network. This becomes problematic if the model encounters inconsistent channel information for the same channel, potentially resulting in reduced performance.

    \item \textbf{System Evolution}:  As direct positioning gains increasing attention in the research community, it's anticipated that findings from current studies might influence future standard revisions. Hence, it's crucial that the initial release of the standard is designed with the flexibility to adapt to these technological advancements.

    \item \textbf{AI/ML model performance degradation after deployment:} 
    It is well understood that AI/ML models can exhibit sub-par performance after deployment~\cite{quinonero2008dataset,chen2023chatgpt}. There are several reasons for this, stemming from changing properties/conditions in the deployment environment that render the available model obsolete to training processes that converge to sub-optimal solutions or training datasets that are not representative of the task/environment at hand. This necessitates the availability of a flexible and robust AI/ML model performance monitoring and management framework. 

    \item \textbf{Service orchestration}: The positioning capabilities introduced in 5G release 16 present a variety of scenarios wherein the position determination can be either UE-driven or network-driven (by the \ac{LMF}), based on uplink, downlink, and, as of release 18, sidelink measurements. 
    
    Transitioning these scenarios to direct positioning poses challenges, as the AI/ML model deployment infrastructure has to account for nuances and variables that differ across these positioning scenarios. Challenges encompass integrating model storage, training, and inference entities, managing model updates and delivery, and ensuring consistent signaling and reporting functionalities across different realization instances.

    \item \textbf{Targeted Data Collection:} 
    Identifying the correct type of measurement data becomes pivotal not just for the system's functionality, but also to consistently meet QoS requirements. 
    
    Collecting data related to exact user locations is fraught not only with privacy but also with technical challenges. Ensuring that data acquisition offers a representative snapshot of diverse reception conditions the positioning system might face is not trivial, especially since the dynamic nature of environments introduces temporal variations, necessitating frequent data refreshes to ensure positioning accuracy. 
    
    Data collection incurs increased costs, especially when it necessitates the involvement of extra entities such as \acp{PRU} for providing ground truth labels, therefore striking a balance between collecting the minimum required amount of~\emph{representative} data becomes imperative.  To add to this, while collecting data from diverse users relying on \ac{RAT} independent sensors, maintaining its integrity becomes essential. Often, within 5G networks, there's limited visibility into the accuracy and quality of the methods used to produce these labels.

\end{itemize}

\renewcommand\bcStyleTitre[1]{\large\textcolor{OliveGreen}{#1}}
\begin{bclogo}[
  couleur=White, 
  arrondi=0,
  logo=\bclampe,
  barre=none,
  noborder=true]{\itshape \textbf{ For the integration of direct positioning, 5G is set to address aspects  spanning operational adaptability, vendor compatibility, system evolution, AI/ML model performance, service orchestration, and targeted data collection.}}
\end{bclogo}


\section{AI/ML model life cycle management}\label{sec:3gpp_sota}

\subsection{Model management complexity}

In an effort to address the challenges analyzed in Section~\ref{sec:challenges}, the cornerstone of the study is an extensive discussion on the properties and requirements of AI/ML model training and deployment~\cite{tr38843}. The first critical question here is whether a single AI/ML direct positioning model, trained on a sufficiently large and diverse dataset from various scenarios, should be deployed or if the aim should be the availability of several smaller/specialized models, e.g., per scenario/area/cell ID, as it is well known to AI/ML practitioners that the largest and more diverse the training dataset is, the higher the generalization capabilities of the trained AI/ML model, but on the other hand, there is a clear trade-off between training dataset size (i.e., generalizability) and model computational and storage complexity.

Another interesting observation is that the use of an AI/ML direct positioning model is not always required. For instance, in \ac{LOS} settings non-AI/ML positioning methods would suffice, while in other scenarios AI/ML assisted positioning could be more suitable.

\renewcommand\bcStyleTitre[1]{\textcolor{RoyalBlue}{#1}}
\begin{bclogo}[
  couleur=Apricot,
  couleurBarre=RoyalBlue,
  arrondi=0,
  logo=\bcbook,
  barre=none,
  noborder=true]{\hspace{0.2cm} \textbf{Example A: Model management}}{\itshape \textcolor{RoyalBlue}{Consider a UE within the environment shown in Figure~\ref{fig:InFexample} that moves from the area in the vicinity of UE-B, towards the area around UE-C, UE-D and finally entering the area where UE-A is located. In this case, when the UE is in the area around UE-B, an AI/ML assisted positioning model needs to be \textbf{activated} and used. As the UE moves towards the area around UE-C, the AI/ML assisted positioning model has to be \textbf{deactivated} and classical positioning methods take over the positioning task. Once the UE enters the area in the vicinity of UE-D, an AI/ML direct positioning model needs to be activated to ensure the best possible positioning performance. As the UE moves towards its final position -- in the area around UE-A, -- the direct same AI/ML direct positioning model is kept active  or this is deactivated and a different, specialized AI/ML direct positioning model is activated for that area.}\footnote{\textcolor{RoyalBlue}{Deactivating a currently active AI/ML model and activating a different AI/ML model is called model \textbf{switching}.}}

  }
\end{bclogo}

\begin{figure*}[!ht]
    \centering
    \includegraphics[width=0.9\textwidth] {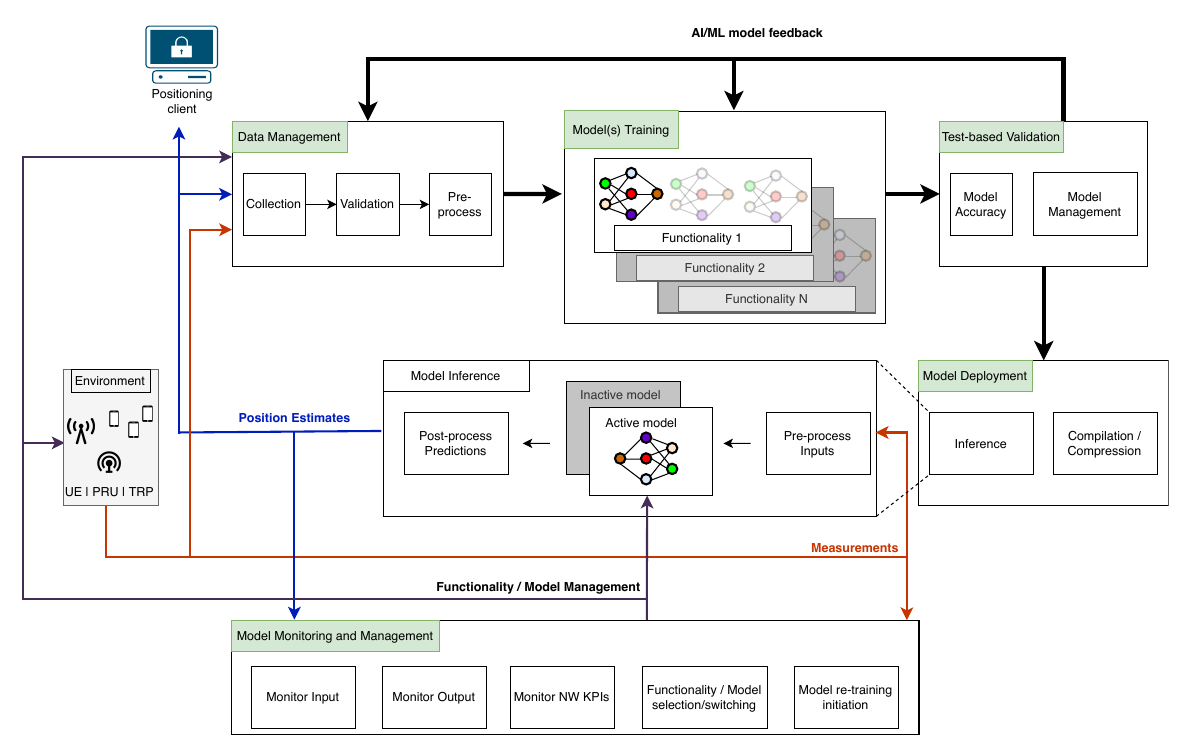}
    \caption{AI/ML direct positioning model life cycle management, as discussed within~\cite{tr38843}  and adapted from~\cite{ashmore2021assuring}.}
    \label{fig:framework}
\end{figure*}

The above example illustrates only a part of the complexity. Another dimension comes from the fact that the AI/ML direct positioning model can be located either at the UE or at the LMF side, as indicated by the direct positioning cases identified within~\cite{tr38843} and shown in Table~\ref{tab:directUsecases}. Here, for UE-side model inference (Case 1), model input data are internally available at UE, while for LMF-side inference (Case 2b and 3b), model input data can be generated by UE (Case 2b) or gNB (Case 3b) and communicated to the LMF. For model training, training data can be generated by UEs, PRUs, gNBs or the LMF.

\begin{table}[!htb]
\caption{ Direct Positioning Cases as defined in \cite{tr38843}.}
\begin{center}
\begin{tabular}{|c|c|c|}
\hline
\textbf{Cases} & \textbf{Inference / Positioning entity}& \textbf{Measurement entity} \\
\hline
Case 1 & UE-side & UE (DL positioning)\\
\hline
Case 2b & LMF-side / LMF & UE (DL positioning)\\
\hline
Case 3b & LMF-side / LMF & gNB (UL positioning)\\
\hline
\end{tabular}
\label{tab:directUsecases}
\end{center}
\end{table}


AI/ML models supporting different sub-use cases (i.e., AI/ML assisted vs AI/ML direct positioning) would most probably require different measurements (or different pre-processing of the same measurements) as model inputs. The same is true for models that support the same sub-use case, but have different capabilities, e.g., models that utilize the full \ac{CIR} information or measurements from all available TRPs usually achieve higher positioning accuracy.

To enable an alignment/coordination mechanism that would ensure that both UE and NW sides are aware of the model data requirements, the concept of \emph{functionality} has been introduced in~\cite{tr38843}. Functionality refers to a feature where AI/ML may be used enabled by configuration(s), where configuration(s) is(are) supported based on conditions indicated by UE capability. The NW-side, taking into account the UE capability reporting, indicates activation, deactivation, fallback, or switching of an AI/ML functionality via 3GPP signalling, in a similar way as in the model management example above.

There may be one or more functionalities defined within an AI/ML-enabled feature, and there can be more than one AI/ML direct positioning models trained per functionality, e.g., models with different structure and size, aiming at different positioning accuracy requirements. Note that in Case 1 (Table~\ref{tab:directUsecases}), where the model resides at the UE, the NW still has control over which functionality is activated. The NW of course has to take into account the UE-supported functionalities (based on its capabilities), but needs not have knowledge on the physical UE-side model. Therefore, the UE can activate/deactivate/switch its model in a transparent way to the NW, under the activated functionality.

To address all these complex requirements and workflows,~\cite{tr38843} discusses the concept of AI/ML model \ac{LCM}, in a similar fashion to non-radio application domains~\cite{ashmore2021assuring}. An illustration of the overall concept and pipeline is shown in Figure~\ref{fig:framework}. Here, the following functional components are identified: 
\begin{itemize}
    \item the offline model training and validation, supported by a flexible data management mechanism;
    \item the online model deployment and inference, enabled by the model monitoring and management entity.
\end{itemize}

\begin{figure*}[!htb]
 \centering
    \includegraphics[width=0.9\textwidth] {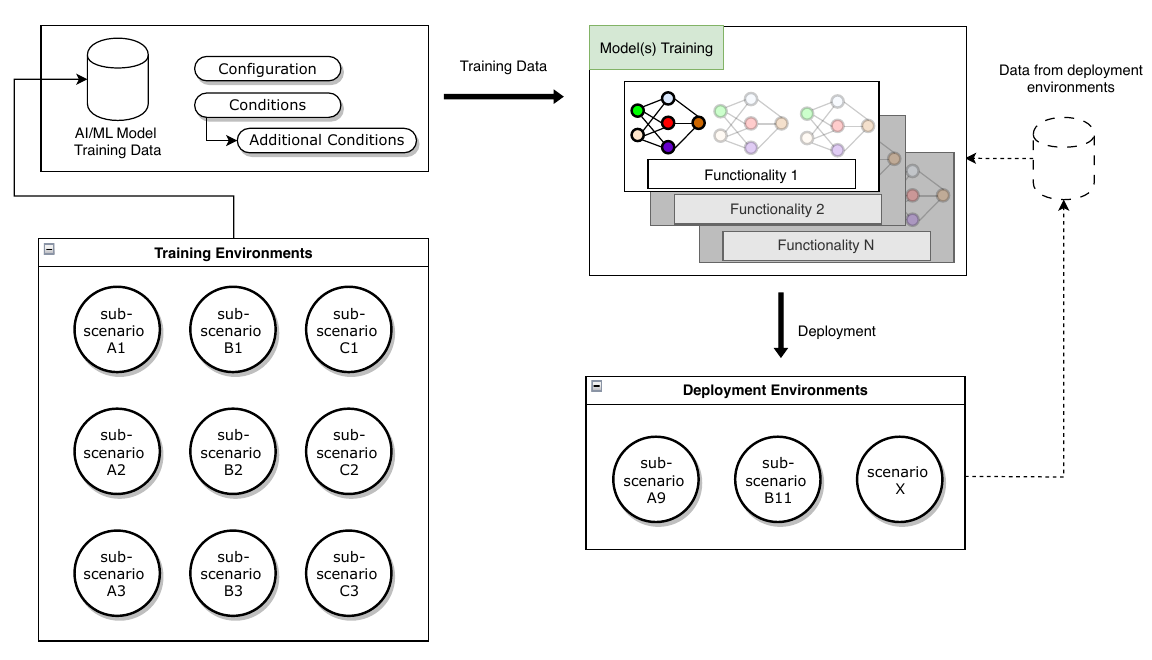}
    \caption{Overview of the semi-supervised learning setup studied within 3GPP~\cite{tr38843}.}
    \label{fig:generalizationTop}
\end{figure*}

\subsection{Data Management} 
This stage is responsible for the acquisition of the required data for the AI/ML direct positioning model training. These data are collected from real or simulated environments and contain all the necessary information for model training, such as measurement (model input), ground truth label (model output), quality indicator associated with the input/output data, time stamp and RS configurations. 

The data need to be validated (e.g., outliers or missing values need to be removed) and pre-processed according to the AI/ML model requirements (e.g., instead of the full \ac{CIR}, \ac{PDP} or \ac{DP} values are extracted or samples with the strongest power are selected, data from specific TRPs are utilized, etc.).

An important point here is the availability of ground truth labels (actual UE positions) to be used for model training. Within~\cite{tr38843} the following options have been identified for ground truth label generation (potentially accompanied by a label quality indicator): i) UE with estimated or known location based on non-NR and/or NR RAT-dependent positioning methods; ii) NW entity based on  NR RAT-dependent or other positioning methods, at least for LMF-based positioning with LMF-side model (Case 2b) and NG-RAN node assisted positioning with LMF-side model (Case 3b).

\subsection{Model training}

Within the recently published TR 38.843~\cite{tr38843}, the AI/ML direct positioning model training discussion revolves around the question of model generalizability. In this direction, Fig.~\ref{fig:generalizationTop} illustrates the overall concept developed and studied within the AI/ML positioning working group. Depending on the selected AI/ML model training approach, a dataset containing training data from a single environment, more than one environments of a specific type (A, B, C) or all the available environments is created and a model is trained. 

The environments have distinct properties, characterized by the following components:
\begin{itemize}
    \item \textit{Conditions:} these are indicated in the UE capability and assist in the AI/ML functionality identification process;
    \item \textit{Additional conditions:} these refer to any aspects that are assumed for the training of the model but are not a part of UE capability for the AI/ML-enabled feature/FG (e.g., UE speed, Doppler, cell ID, etc.);
    \item \textit{Configuration:} here, the type and amount of data to be collected, as well as the properties of the (real or simulated) environment for the data collection are defined. Examples include: number of TRPs, SNR levels, environment geometry, UE speed, configuration of \ac{CIR}/\ac{PDP}/\ac{DP} measurements, etc.
\end{itemize}
Note that some (or all) the environments providing training data could be simulations and not real-word environments~\cite{R1-2309185}. Additionally, reliably labeled data could be scarce in the training scenarios environments. In this case, combining training with semi-supervised~\cite{van2020survey} learning approaches can be utilized. 

Once the trained model is deployed, the question is if it performs adequately for the defined functionality, but under different additional conditions and/or configurations or even in new environments not seen in training. 

The AI/ML research field provides several approaches for developing and training generalizable models in this setting. In almost all of these methods, an amount of labeled data from the target environment is required, with the exact size of this dataset depending on how ``similar'' the source and target environments are. Under this framework, well-studied and widely-used zero-shot or few shot learning~\cite{wang2020generalizing}, meta learning~\cite{hospedales2021meta}, transfer learning~\cite{weiss2016survey,stahlke2022transfer} or fine-tuning~\cite{sharif2014cnn,goodfellow2016deep} algorithms can be applied to ensure generalizability.

\subsection{Model validation} 

Here the trained model accuracy is evaluated, either internally using a dedicated test set either in the form of a RAN4 test (see~\cite{lin2023overview} for more details). In case more than one models are trained and model switching is required during deployment, the model selection/activation/switching mechanism is also evaluated. 

Feedback from this stage (e.g., model success/fail, model over-fitting, etc.) is communicated to the previous stages of the pipeline, to trigger, when necessary, re-training of the model with different parameters and/or even additional data collection.

\subsection{Model Deployment} 

The trained and validated model is then deployed in the (hardware) device to be used for inference. For optimized inference-time and storage complexity, the model needs to be compiled in the device and possibly also undergo a compression/quantization/pruning process~\cite{han2015deep}.

\subsection{Model Inference} 

When the respective functionality and model are activated, measurements (and additional conditions when required) from the environment are provided as model input. The model in turn, provides positioning estimates to the positioning client.

\subsection{Model Monitoring and Management}

Efficient functionality and model management during inference requires a functionality/model monitoring and management layer. The performance of the currently activated functionality/model is constantly monitored and once it degrades, the current model and/or functionality is deactivated and a different one (model and/or functionality) is activated. In the extreme case where the model is deemed no longer valid, e.g., in the case the deployment environment has changed significantly, a new data collection and model (re-) training process can be initiated.

There exist several approaches for AI/ML model performance monitoring, as this is a widely-studied problem in the AI/ML community. The discussion in~\cite{tr38843} follows closely the best practices~\cite{bayram2022concept} and defines performance monitoring metrics based on selected statistics calculated either on the input or the output data, as well as metrics on the deviation between the model predictions and available ground truth labels. Note here that the availability and quality of these labels is governed by the same mechanisms as the ground truth label generation process for the training dataset.

For the different cases defined for AI/ML direct positioning (Table~\ref{tab:directUsecases}), the entity that provides/calculates the monitoring performance metric is the UE or an available PRU for Case 1 and the LMF for Cases 2b and 3b (with LMF-side model).

An additional aspect here, not common in the AI/ML model deployment in other industries/applications, is the concept of functionality/model monitoring and management of inactive functionalities/models. Specifically, when the performance of a model degrades, there could be the option to switch to another (inactive) model and/or functionality. This implies that a mechanism that selects the best possible model/functionality alternative from the set of inactive models/functionalities (or triggers fallback to non-AI/ML positioning approaches) is in place. 
Such a mechanism could be based on hard-coded rules related to the current (additional) conditions in the (radio) environment, on statistics of the input/output data distribution, on experience from past activation/deactivation/switching events under similar conditions, or even from activating candidate models in parallel to the active model and monitoring their performance.

\renewcommand\bcStyleTitre[1]{\large\textcolor{OliveGreen}{#1}}
\begin{bclogo}[
  couleur=White, 
  arrondi=0,
  logo=\bclampe,
  barre=none,
  noborder=true]{\itshape \textbf{ A flexible, scalable, and robust AI/ML model LCM framework is crucial for the viable integration of AI/ML approaches in NR air interface.}}
\end{bclogo}

\section{TR38.843 Direct Positioning Evaluations}\label{sec:eval}

In the preceding sections, we discussed the evolution of positioning technologies in 5G, the challenges they face, and the considerations surrounding AI/ML model life cycle management. Building upon this, we now provide a comprehensive summary and analysis of key evaluations that significantly contributed to the observations and conclusions regarding direct positioning, as detailed in \cite{tr38843}.

Conventional methods tend to be less accurate in environments with a lot of obstructions (NLOS). The specific degree of this inaccuracy is largely determined by additional delay as defined as the \ac{ATOA} Model for the NLOS links in InF environments in \cite{tr38901}. Reports \cite{tr38857} and \cite{tr38843} highlight that resulting errors can range up to several meters.

\subsection{Evaluations assumptions}
Synthetic dataset generated according to the statistical channel models in \cite{tr38901} is used for model training, validation, and testing. The dataset is generated by a system level simulator according to the following scenarios and assumptions: 

\begin{itemize}
    \item TR38.901 \cite{tr38901} outlines various configurations for representing an indoor factory setting. These configurations detail aspects like the size of the hall, the arrangement of TRPs or dropped UEs, and the clutter parameters such as density, height, and size. The evaluation modified the IIoT indoor factory (InF) scenario for FR1 and FR2 as in \cite{tr38857} resulting in the evaluation parameters of Table 6-4.1-1\cite{tr38843}.
     One primary goal for utilizing AI/ML in direct positioning is to evaluate its performance in predominantly NLOS situations. To study the performance under NLOS conditions, the following clutter parameters were prioritized for the common evaluations: clutter density: 60\%, clutter height: 6 meters, clutter size: 2 meters.
     With these settings, in roughly 85\% of the scenarios, none of the TRPs are in direct LOS. Moreover, in over 99\% of situations, one or fewer TRPs are visible in LOS.
    
    \item Spatial consistency for drop-based simulations defines parameter-specific correlation distance in which that the channel generation steps generates spatially consistent clusters. Direct AI/ML model operations rely on the environment-specific mapping of channel observations within a given spot, allowing the model to discern a UE's position. The influence of incorporating spatial consistency on direct positioning was explored in \cite{r1QC}. The findings indicated superior results when the model was trained on channel realizations generated with spatial consistency, in contrast to evaluations conducted without this feature. Additionally,  \cite{tr38901} defines only the decorrelation distance and does not define a specific calibration procedure for spatial correlation (i.e. spatial consistency) of propagation conditions. Therefore, how spatial correlation is implemented can vary. Different interpretations of the LOS/NLOS state's spatial consistency can influence performance results. Factors such as correlations between TRPs, like differences in Angle of Arrival (AoA), are not taken into accounted. For a more comprehensive evaluation, especially when the aim is better generalization across varied clutter settings, the methodology highlighted in TR \cite{SDCpaper} might be a suitable reference.
    
    \item Finally, in the simulations, it's assumed that no tracks are incorporated. This hints that solutions which rely on information from successive snapshots have not been taken into account in the evaluations.
\end{itemize}

\renewcommand\bcStyleTitre[1]{\large\textcolor{OliveGreen}{#1}}
\begin{bclogo}[
  couleur=White, 
  arrondi=0,
  logo=\bclampe,
  barre=none,
  noborder=true]{\itshape \textbf{  Refinements to  the channel models to enable clearer insights into tracking and generalization aspects are needed.}}
\end{bclogo}


\subsection{ Observations on AI/ML Model Performance }

\subsubsection{Multisource Accuracy Evaluations}

In the detailed study outlined in \cite{tr38843}, multiple evaluations were conducted focusing on direct positioning. These evaluations were categorized into five distinct groups. The first group centers on supervised training and testing under identical settings without label errors. The second group delves into evaluations where training and testing environments differ. The third explores the impacts of fine-tuning. The fourth group examines the performance in the presence of label errors, while the fifth incorporates semi-supervised evaluations. 

\begin{figure}[!ht]
    \includegraphics[width=1.0\columnwidth] {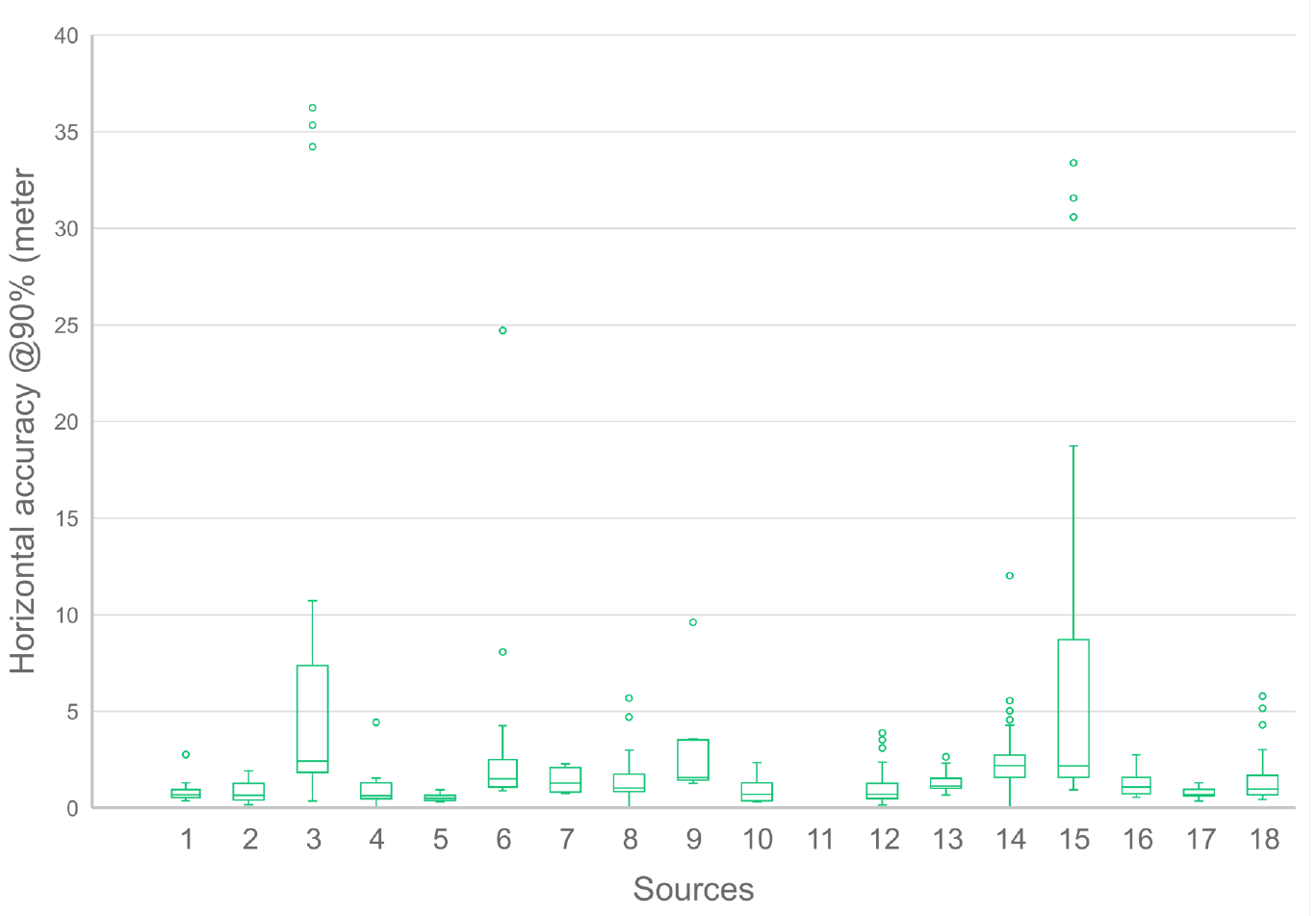}
    \caption{ Results from multiple sources in \cite{tr38843} for direct positioning evaluations, excluding generalization aspects and without label errors.}
    \label{fig:DirectErrALL}
\end{figure}

\begin{figure}[!ht]
    \includegraphics[width=1.0\columnwidth] {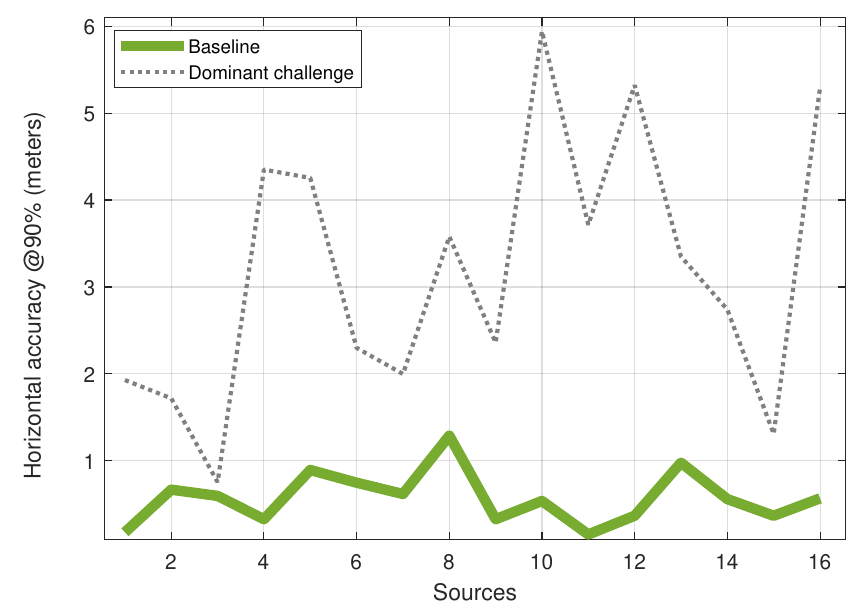}
    \caption{Results from \cite{tr38843} showcase direct positioning evaluations based on multiple sources, using clutter parameters {density 60\%, height 6m, size 2m}. The data highlights baseline performance and the dominant challenge highlighting the highest performance degradation from the identified challenges by each source.}
    \label{fig:DirectErr}
\end{figure}

Fig.~\ref{fig:DirectErrALL} provides a visual representation of results from the first group, sourced from multiple contributors, which specifically exclude generalization factors and label errors. from this set, in Fig.~\ref{fig:DirectErr}, data unrelated to InF clutter parameters ({density 60\%, height 6m, size 2m}) was selectively removed. We also eliminated outliers, which often pointed to situations where the measurement data collection was either too sparse or where the chosen model size was entirely unsuitable. For the scenarios we assessed, the primary performance metric was the 90\% CDF percentile of horizontal accuracy.

The minimal error depicted in Fig.~\ref{fig:DirectErr} primarily illustrates the potential performance of direct positioning when it doesn't face various challenges or particular deployment aspects. Nearly all evaluations consistently reported a positioning accuracy of less than 1 meter. 
The plot showcasing the lowest error is directly associated with the peak error from identical source. This correlation emphasizes that the challenges causing performance degradation can differ substantially depending on the source. Thus, Fig. ~\ref{fig:DirectErr}  offers a general perspective on the potential accuracy decline when certain assumptions or considerations are overlooked.  Understanding both the variance within a single source and the differing performances across multiple sources calls for a more in-depth analysis, aligned with the diverse of the observation and evaluation frameworks.

\renewcommand\bcStyleTitre[1]{\large\textcolor{OliveGreen}{#1}}
\begin{bclogo}[
  couleur=White, 
  arrondi=0,
  logo=\bclampe,
  barre=none,
  noborder=true]{\itshape \textbf{ Direct positioning evaluations show accuracy under one meter. The relationship between the best and worst errors reveals the importance of careful analysis to catch all the challenges.}}
\end{bclogo}

\subsubsection{Time-Domain Model Inputs}

In evaluations, most results for model input utilize time-domain data extracted from the channel response. Time-domain input, as categorized by \cite{tr38843}, falls into three types: \ac{CIR}, \ac{PDP}, and \ac{DP}. CIR uses complex numbers, capturing magnitude and phase, while PDP focuses on power and delay, excluding the phase. In contrast, DP only captures the timing of signal paths.

Choosing the right model input is vital for balancing performance with minimized data collection overhead. The size of this input is paramount in this regard. When evaluations considered consecutive time domain samples and used CIR, PDP, or DP, reducing samples from 256 to 128 had minimal impact on accuracy. Further reduction to 64 or 32 samples decreased accuracy noticeably. In a common approach, focusing on the time domain samples with the strongest power showed that reducing from 256 to 64 samples hardly affected positioning. However, a decrease to 32 or 16 samples impacted performance. Notably, cutting samples from 256 to just 8 significantly compromised accuracy.

As a conclusion on time-domain model input evaluation, it is observed for direct AI/ML positioning, the nature of the model input and the complexity of the AI/ML process directly affect the resulting positioning accuracy. CIR generally surpasses PDP in performance under similar conditions. When other elements, like the model's complexity, becomes relevant, some evaluations observed that PDP and DP might yield better performance than CIR for models with lower complexity.

\subsubsection{Synchronization and Timing Inaccuracies}

In both timing errors and network synchronization error evaluations, the accuracy of AI/ML positioning can degrade when these errors are present and not considered in during training. Additionally, training and inferring the model with such errors, results in improved positioning performance compared to models trained under ideal assumptions \cite{r1Huawei}.

\subsubsection{Influence of SNR variations}

Assessing the influence of SNR on direct AI/ML models reveals that training in optimal SNR conditions enhances positioning accuracy when CIR is the chosen model input. However, when there's an SNR mismatch between training and testing phases, possibly due to alterations in transmit power, the positioning accuracy might degrade. Further evaluations underscored that models trained at higher SNRs adapt more favourably to unknown SNR situations than those trained at lower SNRs.  The positioning error increases significantly when a model trained at one SNR level is tested at a different one.

\subsubsection{Label errors}
While examining the robustness of direct AI/ML positioning to label errors, the study established that positioning remains resilient to specific label errors within a range of 0 to 5 meters. This tolerance to label error is directly related to the positioning accuracy requirements. As accuracy demands become more stringent, the permissible label error should be kept low.
One observation highlighted a potential benefit of utilizing semi-supervised learning. When a limited amount of ideal labelled data was employed, semi-supervised learning showed potential in enhancing positioning accuracy compared to using the same amount of data for supervised learning alone.

 \subsection{Generalization Evaluations }

One primary consideration is how the model performs on datasets it hasn't encountered during training. The evaluations took into account various drops or distinct clutter settings. Different drops can be interpreted as separate industry factories with analogous LOS and NLOS conditions. On the other hand, varying clutter settings can represent different industry factories with distinct LOS and NLOS conditions.

\subsubsection{Fine-tuning}
An AI/ML positioning model was initially trained using one set of data. For fine-tuning, this model was then adapted to a second scenario using a different dataset. It's anticipated that the amount of data needed for this fine-tuning would be significantly less than what's required to train the model from the ground up. After this adaptation, the model's performance was tested to determine if the fine-tuned version could predict locations with a satisfactory degree of accuracy.

  Evaluations were carried for fine-tuning under different drops and different clutter setting. For evaluations, under trained for a given simulation drop (run) and refined with a second drop under the same scenario settings and within a reported sample density.
  
  From the provided evaluation results on different drops, it can be observed that as the dataset size for fine-tuning increases, the positioning error generally decreases. The positioning error regardless of the sample density from the  was larger than 3m for most of the results. Even when the fine-tuning dataset size is increased to 34\% - 50 \% of the full training dataset size. At this range, the error is between 1.22 and 2.70 times of the  training accuracy based on the full fine-tuning dataset.  In a different drop, a new simulation environment is created. Although it shares channel characteristics similar to the training dataset, it lacks environmental correlation with the fine-tuned dataset, this lack of correlation likely explains the observed results.  Conversely, when fine-tuning under different clutter settings, evaluations indicated even poorer results, suggesting that adapting to new clutter conditions is more challenging \cite{r1Nvidia}.

While preliminary evaluations \cite{r1MTK} on time-varying changes have been conducted, a more comprehensive analysis leveraging spatial correlation is essential. This will provide meaningful conclusions regarding the applicability of fine-tuning in environments with time-varying changes.

While only a few evaluations were conducted on the applicability of fine-tuning to address varying SNR values, the results indicated it wasn't suitable. There's potential for fine-tuning to be beneficial in managing different network synchronization and timing errors. Nonetheless, due to the limited scope of evaluations during the study, drawing definitive conclusions isn't feasible.

The report \cite{tr38843} concluded that fine-tuning holds potential if the new deployment scenario is not significantly different from the previous deployment scenario the model was trained for (e.g., 2ns difference in network synchronization error between the previous and the new deployment scenario), fine-tuning a previous model requires a small (e.g., 10\%) training dataset size as compared to training the model from scratch, in order to achieve the similar performance for the new deployment scenario. if the new deployment scenario is significantly different from the previous deployment scenario the model was trained for (e.g., different drops, different clutter parameter, different InF scenarios), fine-tuning a previous model requires similarly large training dataset size as training the model from scratch, in order to achieve the similar performance for the new deployment scenario.

\renewcommand\bcStyleTitre[1]{\large\textcolor{OliveGreen}{#1}}
\begin{bclogo}[
  couleur=White, 
  arrondi=0,
  logo=\bclampe,
  barre=none,
  noborder=true]{\itshape \textbf{ Fine-tuning works best when the deployment scenario is similar to the original training conditions; divergent conditions require new training.}}
\end{bclogo}

\subsubsection{ TRP selection and training strategy}

For the performance evaluation with a diminished count of TRPs, two evaluation strategies were explored. The first method maintained a consistent model input size, but the actual number of TRPs supplying measurements to this input varied, often being less than the input size itself. In the second method, the TRP dimension of the model input matched the count of identified TRPs providing these measurements. For each approach, the TRP set offering measurements for model training could either remain constant or be adjusted dynamically. 

The findings emphasized that recognizing active TRPs is crucial, especially according to the second method with dynamic TRP selection. Without this identification, the model delivered significantly reduced positioning accuracy, sometimes resulting in errors exceeding 10 meters.


\section{Potential Counteractions}\label{sec:Counteractions}

\subsection{Identified potential impact}

The technical report outlines baseline functionalities related to signaling, reporting, and entities to support the operation of the training, inference, monitoring, and management phases. In Section 7.2.4 of \cite{tr38843}, the following aspects and its associated potential specification impact were identified:

\begin{itemize}
    \item \textbf{AI/ML Model Indication:} to enable criteria for selecting and configuring AI/ML models based on their validity conditions, such as the applicable area, scenario, or environment, and their capabilities, such as positioning accuracy and inference latency.
    
    \item \textbf{Signalling, Report/Feedback:} this includes assistance signalling and procedure to facilitate model inference for both UE-side and Network-side model which can comprise new and/or enhancement to existing assistance signalling.

    \item \textbf{Training Data Generation and Collection:} to define different entities and mechanisms for generating the ground truth label and its associated label quality. Particularly, \acp{PRU} are foreseen to play a pivotal role in the direct-positioning framework. PRUs are responsible for generating ground truth labels in various scenarios, ranging from UE-based positioning to NG-RAN node assisted positioning. The implications of integrating PRUs as a key component in this process will further be defined during the work item phase.
    
    \item \textbf{Model Monitoring:}  identifies the metrics and methodologies with potential specification impact employed to evaluate AI/ML models in real-time. This includes how monitoring decisions are indicated between entities and the role of ground truth labels in this process.
    
    \item \textbf{Model Inference-Related Discussions:} the types of existing or new measurements as model inference input and the related measurement reports.      In one detailed recommendation,  CIR and PDP are identified as potential new measurements for Case 2b and 3b where the LMF is the inference entity. 
    
\end{itemize}

\subsection{Selected solutions} 
We highlight key solutions from proposed multiple sources during the study. These solutions, primarily concerning measurement reporting, data collection, and model management, are topics anticipated for discussion in future work.

\subsubsection{Measurement Reporting} 
\begin{figure}[!ht]
    \includegraphics[width=1.0\columnwidth] {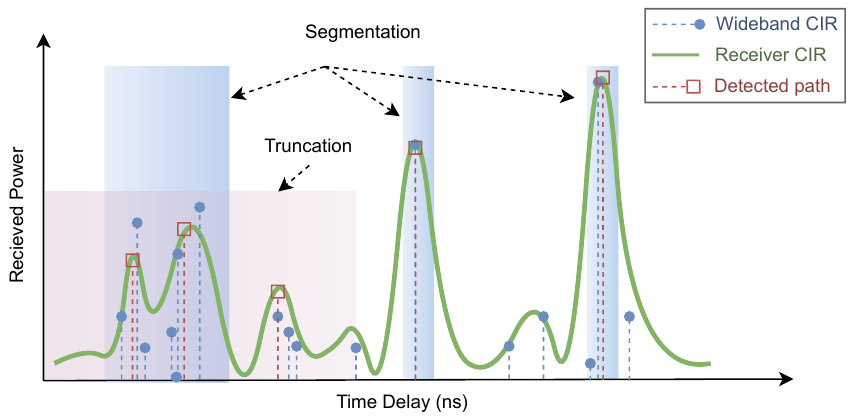}
    \caption{Example of a CIR highlighting truncation, strongest paths and segmentation reporting options.  }
    \label{fig:cirSeg}
\end{figure}

For direct AI/ML positioning using the LMF-side model (Case 2b and 3b), two main challenges arise: Firstly, there's the issue of data size. Reporting the \ac{CIR} can consume significant bandwidth, especially when it includes comprehensive information. The amount of data increases with the inclusion of multiple UE or TRP antennas, multiple snapshots and other parameters in addition to the delay power and phase information, leading to a bulky report. Secondly, there's the challenge of data collection for model training. High-quality labeled data collection, is both resource-intensive and challenging. Additionally, if a future model iteration requires additional parameters, like phase information, a new cycle of costly data collection is triggered
Different measurement reports balancing accuracy and signaling overhead were explored. These reports detail the channel response's timing, power, and phase, considering potential specification modifications like truncation, feature extraction, and sample/path alignment.  In Fig. \ref{fig:cirSeg}, a receiver CIR is illustrated, representing the correlation function and set against the wideband CIR, showing the actual underlying channel paths. The receiver CIR is derived through cyclic correlation, employing a nominal sampling frequency (such as 100MHz, which is the maximum bandwidth for a single carrier in FR1). Several methods optimize report size while targeting accuracy:

\noindent\textbf{Truncation:} By reporting a portion of consecutive CIR samples \cite{R1-2306744}, this approach often focuses on the portion around the first detected path or the strongest path. However, this method might overlook vital data in scenarios with extended delay spreads, likely more evident in un-examined outdoor evaluations.

\noindent\textbf{Strongest Paths:} This method reports paths exceeding a specific power threshold, capturing significant multipath components. However, path selection can be ambiguous. A high detection threshold might overlook important paths, especially with weak LOS signals, while a low one might detect many paths, adding unnecessary overhead.

\noindent\textbf{Segmentation:} Proposed by \cite{R1-2307236}, this method divides the CIR, adjusting to channel characteristics and data collection needs. It offers varied reporting granularity across different segments to dynamically adapt to different environmental contexts and direct- positioning needs. For instance, possible detailed reporting around the first path and selective clusters, but more relaxed for later, undefined clusters can be configured. 

\noindent\textbf{Feature Extraction:}: \cite{R1-2307672} recommends transforming the CIR to extract key features for model training. Using the logarithm of a path's signature, this method efficiently captures essential positioning details for AI/ML models. However, standardizing this approach remains a challenge despite its reported advantages.

\renewcommand\bcStyleTitre[1]{\large\textcolor{OliveGreen}{#1}}
\begin{bclogo}[
  couleur=White, 
  arrondi=0,
  logo=\bclampe,
  barre=none,
  noborder=true]{\itshape \textbf{ The type of selected CIR measurement reports for LMF-side models targets addressing the challenges associated with balancing data complexity and the quality of training data collection.}}
\end{bclogo}

\subsubsection{Data collection}

Having the necessary data density for training and testing/validation is a key challenge for direct positioning. For training data sets evaluation,  a grid distribution, i.e., one training data is collected at the center of one small square grid was defined. Performance variations were identified under different grid assumptions \cite{R1-2306799}. In real-world this can translate to how many reference labels with sufficient quality the data training needs to provide.  Additionally, in practical data scenarios, obtaining a dataset with uniformly distributed ground truths is rare, despite it can be feasible under certain offline data collection (e.g. Raytracing).
\begin{itemize}
    \item \textbf{\ac{IPD}} \cite{R1-2307242} suggests the \ac{IPD} as metric, particularly when data doesn't uniformly spread. IPD measures the proximity between data points, offering insights into the dataset's density and quality. High-density datasets typically have a low IPD, indicating closely packed data points. In contrast, sparser datasets result in higher IPD values. To avoid unnecessary data collection and training, \cite{R1-2307242} applied the derived IPD and compared it against a set threshold to determine the dataset's adequacy. If it doesn't meet the threshold, more data is collected to ensure the machine learning model achieves the intended performance.
    
    \item \textbf{\ac{ACS}}  While the IPD metric primarily concerns the quality and density of data points, \cite{R1-2307236} \ac{ACS} adds a different layer of  related to the information's spatial applicability. ACS, as described in \cite{R1-2307236}, has demonstrated validity within specific regions but may lose its relevance outside those boundaries. This could be tied to factors like varying heights or the evolving characteristics of an area, which can impact the reliability of desired channel information like V-TRP. In positioning, the count of necessary ACS corresponds to a balance of signaling complexity, information availability, and granularity needs.  Thus, while they measure different aspects, ACS can be seen as complementary to IPD, enriching the context and depth of data insights
      
    \item \textbf{Landmarks}  Utilizing direct positioning with landmark, as proposed in \cite{R1-2307236}, can play a pivotal practical role in complex positioning environments when acquiring a ground truth label position is challenging or accompanied by high costs. As reliable reference points with known positions, they offer a robust solution to the challenge of obtaining ground truth labels in such scenarios. Markers identified with sensors like LIDAR or cameras, and the device position relative to these established landmarks can be determined.
 
    \item \textbf{Ground Truth Label Requirements Minimization} There are several methods based on active learning principles~\cite{ren2021survey} developed within the ML community for requesting labeled data only when these data are predicted to improve the model performance, which could be adopted to minimise the acquisition of labeled positioning data. To add to this, AI/ML model training approaches that are robust to noisy label data~\cite{song2022learning} could enable PRU-free, large-scale collection of labeled data with reduced accuracy.

    In the positioning community, approaches based on the idea of channel charting, that aspire to generate a (local or global) chart of measurements that is consistent with the geometry of the radio environment~\cite{stahlke2023indoor}, have enabled reliable positioning and model monitoring with few or no labeled data. Complementary research in a similar direction~\cite{R1-2306744,R1-2304727} enables label-free monitoring. Utilizing displacement information, typically from \ac{IMU} data, emerges as a promising approach, to lessen the reliance on collecting extensive ground truth labels.

    \item \textbf{Proactive model selection and switching}

    Depending on the application, positioning accuracy requirements can vary from 30 $cm$ to 3 $m$~\cite{ts22261}. It is therefore futile to always select and activate a (possibly large/high-latency) model that provides the best possible accuracy, when positioning requirements are more relaxed. What is needed instead is to activate a functionality/model that would fulfil the positioning accuracy constraint, with the least amount of computational, storage and signaling complexity. 

    This need is also there in the functionality/model switching case. First of all, constant switching between  functionalities or specialized models in overlapping areas can lead to similar ``ping-pong'' effects, as observed in handover management settings~\cite{R1-2305197}. There can also be delay-related problems. Models might not be available at the UE and need to downloaded from the NW~\cite{R1-2305197} or, as discussed in~\cite{R1-2304892}, the activation of an AI/ML-enabled functionality/model can induce certain delay when switching from non-AI/ML methods. Additionally,~\cite{R1-2305197} makes the point that switching between functionalities could result in different input requirements for the underlying models, with high-performing models usually requiring mode extensive measurements.

    Based on these observations, it is clear that a \emph{proactive} AI/ML functionality/model management should be able to select for activation the inactive functionality/model that would lead to the smallest (computational, storage, signaling, etc.) mid-term \emph{cost}, while satisfying the posed positioning \emph{performance} requirements.

\end{itemize}

\section{Conclusion}
In our detailed study of AI/ML direct positioning in 3GPP systems, we spotlight the inherent challenges and potential of improving positioning accuracy, especially in complex environments like non-line-of-sight (NLOS) scenarios. In essence, our exploration underscores the value of AI/ML in advancing direct positioning.
 Our analysis, based on TR38.843 study, introduce Life Cycle Management (LCM) for AI/ML positioning which illustrates the need for careful handling of the different process, from their training to real-time monitoring. 
Furthermore, our examination of data synthesis and model fine-tuning highlighted the importance of adaptability in the models. Factors like Signal-to-Noise Ratio (SNR) and synchronization errors can drastically affect performance, underscoring the need for robust and adaptable models.
We present selected solutions designed to counteract inherent challenges. These solutions specifically address areas such as data collection, and model management. We also delved into the complexities of measurement reporting. Techniques like segmentation and truncation help optimize this process, but they come with their own set of challenges.

Looking forward, our analysis show that AI/ML positioning has the potential to be a key positioning application enabler and is likely to be supported and maintained over the future 3GPP releases. 
\label{sec:conclusion}

\bibliographystyle{ieeetr}%
\bibliography{bibliography}%

\vspace{-1mm}








\end{document}